\documentclass[aps,prx,twocolumn,groupedaddress]{revtex4-1}
\usepackage[version=3]{mhchem}
\usepackage{amsmath,amssymb,graphicx,color}
\usepackage{dcolumn}
\usepackage{bm}
\usepackage[mathlines]{lineno}
\usepackage{epsfig}
\usepackage{ulem} 
\usepackage{graphicx}
\usepackage[cp1251]{inputenc}
\usepackage[english]{babel}
\usepackage{epstopdf}

\def\e{\begin{equation}}
\def\f{\end{equation}}
\def\=#1{\overline{\overline #1}}

\def\-#1{{\bf #1}}
\def\.{\cdot}
\def\l#1{\label{eq:#1}}
\def\r#1{(\ref{eq:#1})}

\newcommand{\eps}{\varepsilon}
\renewcommand {\Im}{\mathop\mathrm{Im}\nolimits}
\renewcommand {\Re}{\mathop\mathrm{Re}\nolimits}
\begin{document}

\title{Antenna model of the Purcell effect}
\author{Alexander~E. Krasnok$^{1}$, Alexey~P. Slobozhanyuk$^{1,2}$,
Constantin~R. Simovski$^{1,3}$, Sergei~A. Tretyakov$^{3}$, Alexander N. Poddubny$^{1,4}$, Andrey~E.~Miroshnichenko$^{2}$, Yuri~S.
Kivshar$^{1,2}$, Pavel~A. Belov$^{1}$}
\address{
$^{1}$ITMO University, St.~Petersburg 197101, Russia\\
$^{2}$Nonlinear Physics Center, Research School of Physics and
Engineering, Australian National University, Canberra ACT 0200,
Australia\\
$^{3}$Aalto University, School of Electrical Engineering, Aalto FI-76000, Finland \\
$^{4}$Ioffe Physical-Technical Institute of the Russian Academy of Sciences, St.~Petersburg, 194021, Russia}

\begin{abstract}
The Purcell effect -- the modification of the spontaneous emission rate in presence of resonant cavities or other resonant objects -- is a fundamental effect of quantum electrodynamics. However, a change of the emission rate caused by environment different from free space has a classical counterpart. Not only quantum emitters, but any small antenna tuned to the resonance is an oscillator with radiative losses, and the influence of the environment on its radiation can be understood and measured in terms of the antenna radiation resistance. We present a general approach which is applicable to measurements of the Purcell factor for radio antennas and to calculations of these factors for quantum emitters. Our methodology is suitable for calculation and measurement of both electric and magnetic Purcell factors, it is versatile and applies to various frequency ranges. The approach is illustrated by a general equivalent scheme and allows the Purcell factor to be expressed through the continious radiation of a small antenna in presence of the environment.
\end{abstract}

\maketitle
\tableofcontents

\section{Introduction}
\begin{figure*}
\centering \epsfig{file=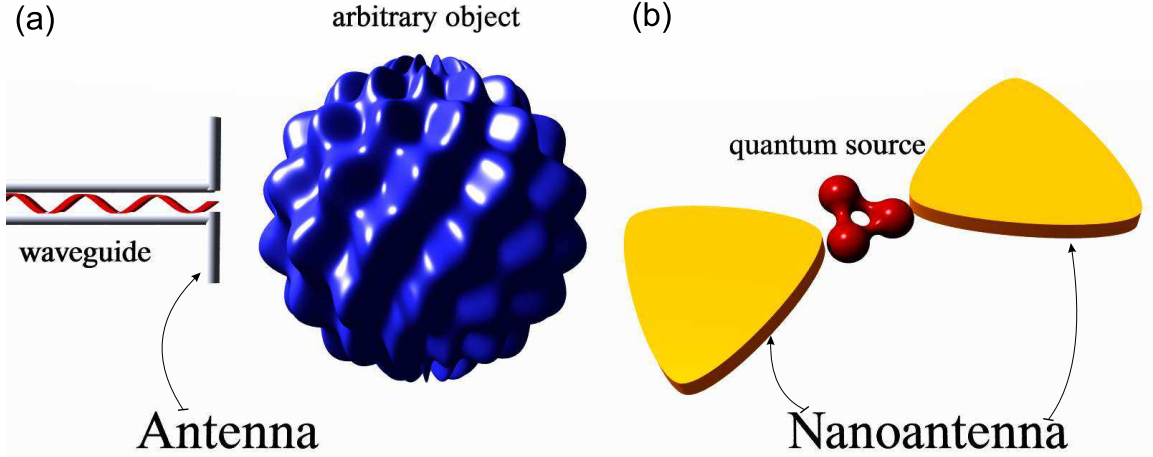,width=0.85\textwidth}
\caption{The classical (a) and quantum (b) realizations of the Purcell effect.} \label{FigGeom1_1}
\end{figure*}

The Purcell effect is defined as a modification of the spontaneous emission lifetime of a quantum source induced by the interaction with its environment~\cite{Purcell_46, Maksymov_PRL_2013, Moerner_bowtie_09, Vahala_2003, Asano_2007, Hu_NAT_2012, Waks_2013, Carminati_PRL_14, Iorsh_2013, Cano_2013}. This modification is significant if the environment is a resonator tuned to the emission frequency. This effect has been first pointed out by E.M.~Purcell~\cite{Purcell_46} in 1946 in the context of nuclear magnetic resonance experiments. At present, this effect is widely used in microcavity light-emitting devices~\cite{kavbamalas,Fainman_2010, Fainman_2013}, in single-molecule optical microscopy~\cite{Sandoghdar_Nature_00, Koenderink_PRL_11, Cosa_2013, Wilde_2014}, in microscopy of single NV centers in nanodiamonds~\cite{Vamivakas_NanoLetters_13}, of Eu$^{3+}$-doped nanocrystals~\cite{Carminati_PRL_14} and for visualization of biological processes with participation of large molecules, such as DNA~\cite{Tinnefeld_Science_2012}. An overview of nanosensing applications of the Purcell effect is presented in~\cite{Kumar2013}.

We start from a brief overview of several equivalent definitions of the Purcell factor -- the value which describes this effect quantitatively -- and the existing approaches to its theoretical and experimental evaluation. First, let us point out that the Purcell effect is based on the quantum electrodynamics concept of weak coupling of an emitter and a resonating object (nanoantenna ~\cite{Kumar2013}, or optical cavity~\cite{kavbamalas,Fainman_2010, Fainman_2013}). The weak and strong coupling regimes~\cite{Novotny_Hecht_book, Khitrova2006} can be distinguished by comparing the so-called emitter-field coupling constant $\chi=[|\mathbf{d}|^2\omega_{0}/(2\hbar\varepsilon_{0}V)]^{1/2}$ with the decay rate of the photon in a cavity $\gamma$ and the nonradiative decay rate of the excited state $\gamma_{\rm dis}$. Here, $\omega_{0}$ and $\mathbf{d}=e\langle2|\mathbf{r}|1\rangle$ are the frequency of the excited-to-ground state transition ($2\to 1$) and its dipole moment (matrix element), respectively, $e$ is the electron charge, $V$ is the effective volume of the resonator mode, $\varepsilon_{0}$ is the vacuum permittivity. We use the SI units, the result in the CGS units can be obtained by replacing $\eps_{0}$ by $1/(4\pi)$.

In the weak-coupling regime, when $\chi\ll\gamma,\gamma_{\rm dis}$, the hybridization of the quantum emitter and the resonator eigenstates is weak. Therefore the frequency $\omega_0$ of the spontaneous emission is not modified by the resonator, and the interaction only leads to a modification of the decay rate. The dipole moment of the optical transition $\mathbf{d}$ and its classical dipole moment $\mathbf{d}_1$ keep unperturbed and $\mathbf{d}_1=2\mathbf{d}$ (see for example Ref.~\cite{Book_AtomMol}, pp.250--251). The ratio of the decay rate $\gamma$ in the vicinity of the resonator to the decay rate of the same emitter in free space $\gamma_0$ can be written as~\cite{Novotny_Hecht_book}:
\e\label{PurcellQ} F\equiv\frac{\gamma}{\gamma_{0}} =
1+\frac{6\pi\varepsilon_0}{|\mathbf{d}_1|^2}\frac{1}{q^3}\Im [\mathbf{d}_1^\ast\cdot \mathbf{E}_{{\rm s}}(\mathbf{r}_{\rm d})],\f
where $\gamma_0=\omega_{0}^3|\mathbf{d}_1|^2/(12\pi\varepsilon_{0}\hbar c^3)=\omega_{0}^3|\mathbf{d}|^2/(3\pi\varepsilon_{0}\hbar c^3)$~\cite{Novotny_Hecht_book}, $q=\omega/c$ is wavenumber in free space, $\mathbf{E}_{\rm s}(\mathbf{r}_{\rm d})$ is the magnitude of an electric field of the quantum source dipole $\mathbf{d}_1$ oscillating at the frequency $\omega_{0}$, scattered from an inhomogeneous environment and evaluated at the source origin $\mathbf{r}_{\rm d}$. The quantity $F$ is called the Purcell factor.
According to Eq.(\ref{PurcellQ}), the magnitude of the Purcell factor does not depend on the magnitude of the transition dipole moment $\mathbf{d}$, because the scattered field value is directly proportional to the dipole moment. For the purpose of this paper it is important to note, that equation~\eqref{PurcellQ} can be applied not only to the cavities or nanoantennas but to an arbitrary electromagnetic environment of the emitter different from free space~\cite{Kumar2013}.  Moreover, the concept of Purcell's factor can be extended to optical emitters which cannot be modeled as a point electric dipole~\cite{Lodahl2011, Kavokin1991}. The Purcell factor can be also understood in terms of the local density of photonic states modified by the presence of the object~\cite{lagendijk_review}.

The above expression for $\gamma_0$ does not take into account the non-radiative decay (it is assumed that $\gamma_{\rm dis}\ll \gamma_0$) and results from the standard formula for the power radiated by a Hertzian dipole $\mathbf{d}_1$ at frequency $\omega_0$:
\e
P_{0,\rm rad}=\frac{\omega_0^4d_1^2}{12\pi \varepsilon_{0}c^3}=\frac{\omega_0^4d^2}{3\pi \varepsilon_{0}c^3},\l{Her}\f
namely,
\e \gamma_0= \frac{P_{0,\rm rad}}{\hbar\omega_0}\f
is the ratio of $P_{0,\rm rad}$ to the photon energy. In the weak coupling regime, the environment modifies only the radiated (far-zone) power and the dissipation of power in the volume outside of the emitter. Thus, the decay factor modified by the environment can be written as
\e \gamma=\frac{P_{\rm rad}+P_{\rm nonrad}}{\hbar\omega_0},\f
and the Purcell factor can be expressed also as

\e \l{expr5} F\equiv\frac{\gamma}{\gamma_{0}} =
\frac{P_{\rm rad}+P_{\rm nonrad}}{P_{\rm 0,rad}}\equiv F_{\rm rad }+F_{\rm nonrad }.
\f
Here, $P_{\rm rad}$ is the power radiated in the far zone (enhanced by the environment) and $P_{\rm nonrad}$ is the power dissipated in the environment.

If the electromagnetic environment is lossless, the last term vanishes and the Purcell factor describes the change of the total radiated power $P_{\rm rad}$ at the frequency of the emitter:
\e \label{PurcellRad} F\equiv\frac{\gamma}{\gamma_{0}} =
\frac{P_{\rm rad}}{P_{\rm 0,rad}},\f
where the index 0 still means the corresponding value for the same emitter in free space. If the emitter is located in a lossy medium (perhaps inhomogeneous) the Purcell factor Eq.~\eqref{PurcellQ} has two contributions: that corresponding to the far-field emission and that corresponding to the Joule losses in the environment~\cite{barnett1996}. When this environment can be described by position-dependent dielectric constant $ \varepsilon(\mathbf r')$ the Joule loss contribution into Purcell's factor can be presented as~\cite{Welsch2006}:
\begin{equation}
F_{\rm nonrad }=\frac{6\pi\varepsilon_0}{q^{3}|\mathbf{d}_1|^2} \int {\rm d}^{3} r' \Im [\varepsilon(\mathbf r')]\mathbf
|\mathbf E(\mathbf r')|^{2}\:,
\end{equation}
where $\mathbf E(\mathbf r')$ is the total field produced by the dipole ${\bm d}_1$ at the point $\mathbf r'$ which is integrated over the surrounding space.

Here, one may introduce the radiation efficiency of the quantum source $\xi$ in the same way as it is done in the antenna theory~\cite{Balanis}: $\xi\equiv F_{\rm rad}/F= (F-F_{\rm nonrad})/F$. The total quantum yield of the emitter $Q$ is determined by the competition between the far-field radiation, the Joule losses, and the internal non-radiative losses of the emitter $\gamma_{\rm dis}$:
\begin{equation}
Q=\frac{\gamma_{0}F_{\rm rad}}{\gamma_{0}F+\gamma_{\rm dis}}\:.\label{eq:yield}
\end{equation}

Note, that in formula (\ref{PurcellRad}) we assumed the decay rate of the emitter in free space to be equal $\gamma_0$, i.e. neglected the non-radiative losses inside the emitter. This can be a realistic approximation for many quantum dots and fluorescent dye molecules (e.g. in~\cite{new} $\gamma_{\rm dis}$ and $\gamma_{0}$ were separately measured for nanocrystal quantum dots and it was shown that
$\gamma_{\rm dis}\ll \gamma_{0}$).

For quantum emitters the total Purcell factor is measured either directly by evaluating the speedup of the time-resolved photoluminescence~\cite{Iorsh_2013} or indirectly, for example, using the Raman spectroscopy~\cite{Boucaud_2010}. High values of $F$ can be achieved with nanoantennas -- resonant devices that effectively convert the near field of quantum sources to propagating optical radiation~\cite{NovotnyAntennasForLight, Krasnok_UFN_2013}. This transformation is carried out by means of impedance matching between the quantum source and the nanoantenna~\cite{Raschke_2012, Alu_dip_PRL_08, Hecht_009, Marquier_PRL_2010}. Another possibility to attain large values of the Purcell factor Eq.~\eqref{PurcellQ} is provided by hyperbolic metamaterials (see the review~\cite{Iorsh_2013}).

In this paper, we look at the Purcell effect in the broader context and study it within the classical framework. While the question whether the spontaneous decay itself is a truly quantum~\cite{ScullyZubairy} or not a purely quantum phenomenon~\cite{GinzburgUFN, Ghiner_2000} is still under debate, the {\itshape modification} of the spontaneous decay in a medium can be definitely considered classically. Indeed, the electric field $\mathbf E_{\rm d}(\mathbf r_{\rm d})$ of the emitter ("antenna" in the Fig.\ref{FigGeom1_1}(a) or "quantum source" in the Fig.\ref{FigGeom1_1}(b)), entering the expression of the Purcell factor Eq.~\eqref{PurcellQ} is a well-defined quantity in optics as well as in classical physics and antenna engineering. Developing this concept, in Section~\ref{Method} we propose a new methodology of calculation and measurement of the electric and magnetic Purcell factors through the input impedance of an equivalent small antenna, generalizing in Subsection~\ref{GeneralMethod} the results of Refs.~\cite{Kumar2013, Marquier_PRL_2010, Slobozhanyuk2014magnetic, Krasnok_superdir_APL} and suggesting in Subsection~\ref{Scheme} an equivalent scheme of nanoantenna's Purcell effect suitable for the direct calculation of involved impedances. As an illustration, in Section~\ref{MethodOptics} we show how our method can be applied to the analytical and numerical calculations of the Purcell factor of nanoantennas for quantum emitters. Our approach naturally results in a method for calculations and direct measurements of Purcell factors through the input impedance of arbitrary \textit{electric and magnetic dipole antennas}. Finally, we present an example of the method application for radio frequencies: Section~\ref{Methodexper} presents an experimental verification of the proposed method for {microwave electric and magnetic dipole antennas} located above a metal mirror.

\section{Retrieval of the Purcell factor through the input impedance}
\label{Method}

\subsection{General Method}
\label{GeneralMethod}
Consider an arbitrary radiating electric dipole with the moment $\-d$ in presence of an arbitrary passive object. Let us attribute No 1 to the dipole and No 2 to the object. The total electric field created by the dipole 1 at its origin $\-E_1({\-r}_d)$ can be decomposed into two parts $\-E_{1}({\-r}_d)=\-E_{11}({\-r}_d)+\-E_{12}({\-r}_d)$, where $\-E_{11}({\-r}_d)$ is the field created by the dipole 1 in the absence of the object 2 and $\-E_{12}({\-r}_d)\equiv \-E_{\rm s}({\-r}_d)$ is the field scattered by the object. The total power delivered by the radiating particle to the environment reads
\e P= P_{\rm{rad}}+P_{\rm nonrad} = -\frac{1}{2}\int_V\Re \left[\mathbf j^{\ast}_1(\mathbf{r})\cdot\mathbf{E}_{1}(\mathbf{r})\right]{\rm d}V, \f
where $V$ is the volume of the radiating dipole and $\mathbf j^{\ast}_1$ is the electric current density in that volume. It splits into two parts $P=P_{11}+P_{12}$, where $P_{11}$ is the power radiated by the same dipole in the absence of object 2 (the same as $P_{0,\rm rad}$ as above), and
\e P_{{12}} = -\frac{1}{2}\Re \left[\mathbf{E}_{12}(\mathbf{r}_d)\cdot \int_V\mathbf j^{\ast}_1(\mathbf{r})\, dV \right]. \l{Pur}\f
Here we assume that the radiating dipole 1 has a sufficiently small volume, such that the spatial variation of field $\mathbf{E}_{12}$ over $V$ can be neglected. Since the classical electric dipole moment is defined through the electric current density ${\-j}_1$ as
\e
{\-d}_1={1\over j\omega}\int_V\mathbf{j}_1(\mathbf{r})\, dV
\f
for the time dependence in the form $\exp(j\omega t)$, formula \r{Pur} can be rewritten as
\e
P_{12}=-\frac{\omega}{2}\Im \left[{\-d}_1^{\ast}\cdot{\-E}_{12}(\mathbf{r}_{\rm d})\right].
\l{Pur1}\f
Formula \r{expr5} after substitution of \r{Her} for $P_{11}$ gives for the Purcell factor
$F$:
\e
F={P_{11}+P_{12}\over P_{11}}=1+\frac{6\pi\varepsilon_0}{|\mathbf{d}_1|^2}\frac{1}{q^3}\Im [\mathbf{d}_1^\ast\cdot \mathbf{E}_{{\rm 12}}(\mathbf{r}_{\rm d})]
\f
i.e. the known result Eq.(\ref{PurcellQ}), which is definitely applicable to both classical and quantum emitters (weakly coupled to an arbitrary object).

The last result can be rewritten in terms of the input impedances and in terms of the Green function. First, by definition of radiation resistance we have:
\e
P=P_{11}+ P_{12}=|I_1|^2R_{\rm rad}=|I_1|^2(R_{0,\rm rad}+R_{12}).
\l{new}\f
Here the effective current $I_1$ referred to the origin ${\-r}_d$ is related with the dipole moment as $I_1=j\omega d_1/l_1$, $l_1$ is the effective length of the dipole 1.
The radiation resistance of an optically small particle with the effective length $l$ reads as~\cite{Balanis}:
\e R_{0,\rm rad}={\eta\over 6\pi}(kl)^2,\l{R_rad}\f
where $\eta=\sqrt{\mu_0/\varepsilon_0\varepsilon_h}$ and $k=q\sqrt{\varepsilon_h}$ are the wave impedance and the wave number of the host medium, respectively.
The additional (mutual) resistance $R_{12}=\Re{Z_{12}}$ caused by the field scattered from the radiation-enhancing object ${\-E}_{12}$ can be found separately. This is an interesting and relevant problem which will be studied in the next subsection.

However, formula \r{new} rewritten as $F={P/P_{11}}=R_{\rm rad}/R_{0,\rm rad}$ may already serve as a practical alternative to the commonly used expression (\ref{PurcellQ}). From the general theory of antennas it is well known that the input resistance of a short dipole equals the radiation resistance, when the dissipative losses inside the antenna are neglected~\cite{Balanis}. Thus, if our emitter is low-loss ($\gamma\gg \gamma_{\rm dis}$), we can write an equivalent relation for the Purcell factor of an arbitrary object (inhomogeneous environment) for a low-loss emitter (does not matter -- quantum or classical):
\e
F=\frac{R_{\rm in}}{R_{0,\rm in}}\equiv \frac{\Re{Z_{\rm in}}}{\Re{Z_{0,\rm in}}}.
\l{new1}\f
In some situations it may be easier to measure or calculate the input impedances ($Z_{0,\rm in}$ and $Z_{\rm in}$) of the emitter 1 in the absence and presence of object 2 than to accurately find the scattered field. Then formula \r{new1} allows the Purcell factor through the real parts of these impedances. This factor for a dipole emitter is determined by the modification of the resistive part of its input impedance. Non-radiative losses in this formula are present in $R_{\rm in}$ since the additional resistance $R_{12}$ is not purely radiative. Mutual coupling
effectively brings the losses of the object 2 into emitter 1.

Let us now show that formula \r{new1} fits another known representation of the Purcell's factor -- through Green's function~\cite{Lenac}.
The electric field produced by a dipole $\mathbf{d}_1$ stretched along the $z$-axis is related to the dyadic Green function of an inhomogeneous environment $\hat{G}(\mathbf r,\mathbf r_d,\omega)$ as follows~\cite{Novotny_Hecht_book}:
\e \label{FG} \mathbf{E}(\mathbf{r})=\frac{k^2}{\varepsilon_0\varepsilon_h} \hat{G}_{zz}(\mathbf{r},\mathbf{r}_d,\omega)\mathbf{d}_1. \f
In order to relate the Green function to the input impedance $Z_{\rm in}$ of our dipole 1 we use the Brillouin method of induced electromotive forces (IEMF)~\cite{Balanis}:
\e Z_{\rm in}=\frac{1}{I_1^2}\int_V {\-E}_{1}(\-r)\cdot \-j_{1}(\-r)dV. \f
The value in the numerator is called IEMF in radio science, and $I_1$ in the denominator is the current though the central cross section of the dipole. In the short antenna approximation $kl_1\rightarrow 0$ the result reads as~\cite{Balanis}:
\e Z_{\rm in}=\frac{E_{1}({\-r}_d)l_1}{2I_{1}}.\f
This expression establishes a relationship between the magnitude of the input impedance of a short dipole and the value of the \textit{total} (not only scattered) electric field $\-E_1=E_1\-z_0$ at the dipole origin.
Now, recall the definition of the Green function Eq.(\ref{FG}) and using the relation $I_1l_1=j\omega d_1$ we obtain an expression that links the Green function with the input impedance of the dipole 1:
\e G_{zz}(0,0,\omega)=\frac{4j\omega\varepsilon_{0}}{l_1^2k^2}Z_{\rm in} \l{ggg}\f
Now we may rewrite \r{new1} in form
\e \label{PurcellImped} F=\frac{R_{\rm in}}{R_{\rm in}^{(0)}}=\frac{\Im G_{zz}(0,0,\omega)}{\Im G_{zz}^{(0)}(0,0,\omega)}. \f
This expression is equivalent to formula (2.4) from~\cite{Lenac}.

Thus, it is possible to find the Purcell factor either using the standard techniques, such as Eqs.~(\ref{PurcellQ}), (\ref{PurcellImped}), or using
formula \r{new1}, in terms of input resistances.

\subsection{Equivalent Circuit for Finding the Purcell Factor}
\label{Scheme}

Here we explain how to find the input resistance $R_{\rm in}$ of an optical (e.g. fluorescent) emitter in the presence of an optically small resonator. Such resonators, called \textit{nanoantennas,} are used to enhance the spontaneous emission of isolated quantum emitters (see e.g. in~\cite{NovotnyAntennasForLight, Hecht_009, Kumar2013}). Though this treatment is targeted to quantum emitters, our consideration is fully classical and based on the concept of electromagnetically coupled oscillators. Therefore, it is relevant to illustrate our approach by equivalent circuits.
\begin{figure*}
\includegraphics[width=1.1\columnwidth]{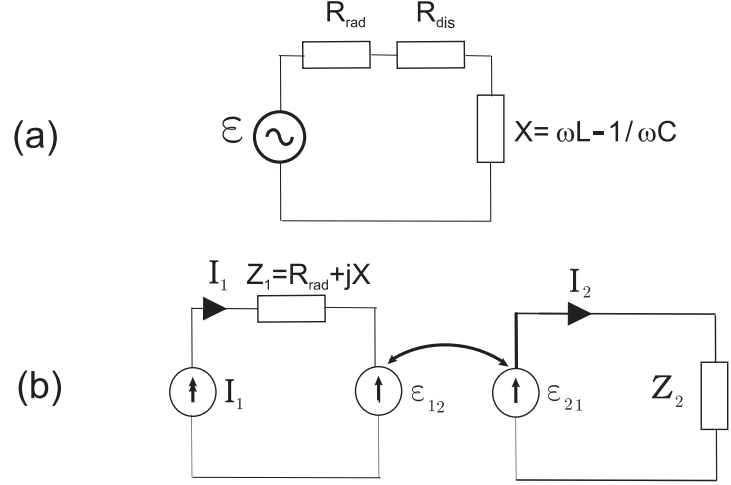}
\caption{(a) An equivalent scheme of a resonant dipole scatterer. (b) Equivalent schemes of an emitter and a nanoantenna in terms of induced electromotive forces.}
\label{fig1}
\end{figure*}

For instance, we notice that in spite of physical differences between a quantum emitter and a nanoantenna (which is a classical resonant scatterer),
both these objects in the absence of tunneling effects interact purely electromagnetically, and their coupling is governed by Maxwell's equations. Therefore, both of them can be described in terms of resonant $RLC$-circuits. An attempt to build such schemes was done in work~\cite{Marquier_PRL_2010}, however without practical results. In the present paper we introduce an alternative equivalent circuit for radiating systems comprising an optical emitter and a nanoantenna. This circuit illustrates a simple algorithm for calculating the additional term $R_{12}$ entering the input resistance $R_{\rm in}$ in presence of object 2. This term is called mutual resistance $R_{m}\equiv R_{12}$.

First, we recall the well-known circuit model of an optically small dipole scatterer excited by an external electric field $\-E=\-z_0 E$ (see e.g. in~\cite{Jackson}). The current, induced in a short dipole antenna of effective length $l$ reads as $I=El/Z$, where $Z=R_{\rm rad}+R_{\rm dis}+jX$ is the total impedance of the particle, see Fig.~\ref{fig1}(a). Here we have split $\Re (Z)$ onto the radiation resistance $R_{\rm rad}$, and dissipation resistance $R_{\rm dis}$. Since the induced dipole moment equals $d_{\rm ind}=Il/j\omega$ (assume for simplicity that $\-d_{\rm ind} =\-z_0 d_{\rm ind}$ that holds for a spherical particle for any polarization and for an ellipsoidal one polarized along one of its axes), the inverse polarizability $\alpha^{-1}\equiv E/d_{\rm ind}$ reads as
\e {1\over \alpha}={1\over l^2}\left[j\omega (R_{\rm rad}+ R_{\rm dis}) -\omega X\right]. \l{alpha_inv}\f
Substituting \r{R_rad} into \r{alpha_inv}, we find
\e {1\over \alpha}=j{k^3\over 6\pi\varepsilon_0\varepsilon_h}+ j{\omega R_{\rm dis}\over l^2} -{\omega X \over l^2}. \l{alpha_inv_k3} \f
This is the well-know formula for the polarizability of a lossy dipole scatterer which is applicable to both quantum emitter and nanoantenna. However, in this paper we neglect the induced part of the dipole moment of the quantum emitter as well as the hybridization of its states. In the weak coupling regime $\-d_1(\omega_0)=2\-d$. However, the emission spectrum has the Lorentzian shape~\cite{Novotny_Hecht_book}, and this means that we have to consider the polarization of the nanoantenna at any frequency $\omega$. So, we use the polarization model \r{alpha_inv_k3} for the nanoantenna. Note, that using \r{alpha_inv_k3} it is easy to find the general limitations on the absorbing and scattering cross sections of the nanoantenna (see e.g. a review~\cite{Tret}). The equivalent circuit of the nanoantenna is shown on Fig.~\ref{fig1}(a) and it contains an IEMF ${\cal E}=El$ loaded by a series connection of the antenna radiation resistance $R_{\rm rad}$, the dissipation resistance $R_{\rm dis}$, capacitive impedance $1/j\omega C$ and inductive one $j\omega L$. This series connection corresponds to the Lorentzian model of the scatterer's dispersion:
\e {1\over \alpha}={1\over \alpha_0}(\omega_0^2-\omega^2+j\omega\Gamma_{\rm dis})+j{k^3\over 6\pi\varepsilon_0\varepsilon_h}.\l{Lor}\f
Comparing \r{alpha_inv_k3} and \r{Lor} we can relate the equivalent parameters with the corresponding parameters $\alpha_0$, $\omega_0$ and $\Gamma_{\rm dis}$ of the Lorentzian model:
\e R_{\rm dis}=l^2{\Gamma_{\rm dis}\over \alpha_0},\qquad
L={l^2\over \alpha_0}, \quad C={\alpha_0\over l^2\omega_0^2}\l{super}\f
Obviously, for the resonance frequency we have $\omega_0^2=1/(LC)$.
\begin{figure*}
\includegraphics[width=1.1\columnwidth]{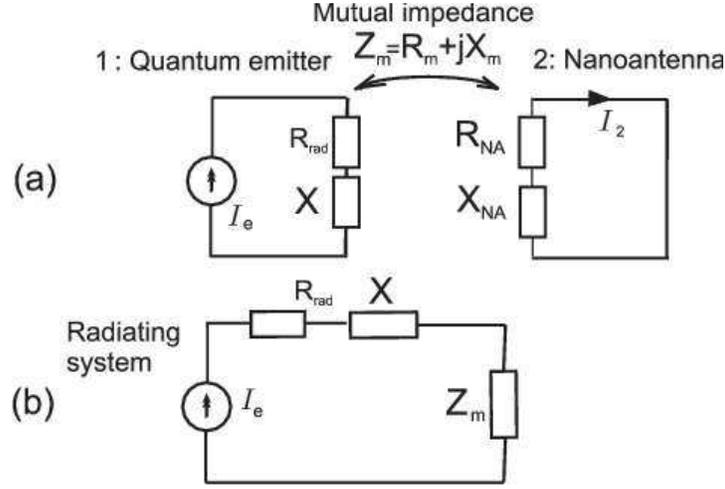}
\caption{(a) Equivalent schemes of an optical emitter and a nanoantenna in terms of mutual impedance. (b) An equivalent scheme of an emitter with the mutual impedance added by the nanoantenna.}
\label{fig2}
\end{figure*}

For the emitter we also start from the general circuit model Fig.~\ref{fig1}(a). This equivalent circuit corrects and replaces an incorrect scheme suggested in~\cite{Marquier_PRL_2010} (Fig.~3(a)).
However, in the approximation of weak coupling as above we assume that the electric dipole moment $\-d_1=d_1{\-z}_0$ is fixed at any frequency corresponding to the emission spectrum. Because the dipole moment is related to the effective current of the emitter $I_e= j\omega d_1/l_1$, the equivalent circuit in Fig.~\ref{fig1}(b) comprising both emitter and nanoantenna is driven by a fixed current source $I_1\equiv I_e$. The replacement of the circuit driven by the EMF $\cal E$ in Fig.~\ref{fig1}(a) by the circuit shown in Fig.~\ref{fig1}(b) which is driven by the current generator is granted by the well-known equivalent generator theorem. In Fig.~\ref{fig1}(b) we neglect the dissipation in the quantum source since the main mechanism of the decay rate is radiative ($R_{\rm dis}\ll R_{\rm rad}$). Here, for simplicity of notations, $R_{\rm rad}$ denotes the proper radiation resistance of the emitter denoted above as $R_{0,\rm rad}$.

The IEMF describing the mutual coupling of nanoobjects 1 and 2 in Fig.~\ref{fig1}(b) can be replaced by mutual impedance $Z_m$, whose real part comprises an additional radiation resistance arising in the emitter and responsible for the Purcell factor. The corresponding modification of the equivalent scheme from the mutually induced EMF to the mutual impedance can be accompanied by following speculations. The emitter induces the IEMF ${\cal E}_{21}=E_{21}l_2$ in nanoantenna 2, where the field $E_{21}$ is that produced by the emitter at the center of the nanoantenna ${\-r}_2$. This field can be written in form ${\-E}_{21}={\-z}_0A_{ee}d_1$, where $A_{ee}$ is the electric field of a unit electric dipole with the origin at ${\-r}_1\equiv{\-r}_d$ evaluated at ${\-r}_2$. In the case of symmetric mutual location of objects 1 and 2 the value $A_{ee}$ is scalar. This IEMF is related with the current induced in the nanoantenna as $I_2={\cal E}_{21}/Z_2$, where $Z_2$ is the impedance of the nanoantenna. The dipole moment of the latter $d_2=I_2l_2/j\omega={\cal E}_{21}l_2/j\omega Z_2$ generates the scattered field $E_{12}$ and the IEMF ${\cal E}_{12}=E_{12}l_1$ arises in the quantum emitter. Due to the reciprocity we may express ${\-E}_{12}$ through the same coefficient $A_{ee}$:
\e
{\-E}_{12}={\-z}_0A_{ee}d_2={A_{ee}^2d_1l_1l_2^2 \over j\omega Z_2}.\l{inter}\f
Since the current in the emitter is fixed, $I_1=j\omega d_1/l_1=I_e$, the IEMF ${\cal E}_{12}=E_{12}l_1$ is equivalent to the mutual impedance $Z_m=-{{\cal E}_{12}/I_1}$ in accordance to the equivalent generator theorem. The minus sign in the relation $Z_m=-{{\cal E}_{12}/I_1}$ appears because the IEMF ${\cal E}_{12}$ is directed oppositely to the driving current $I_1$ (Fig.~\ref{fig1}(b)). Thus, in our final equivalent scheme Fig.~\ref{fig2}(a) the IEMF ${\cal E}_{12}$ is replaced by the mutual impedance $Z_{m}$ describing the contribution of the nanoantenna into the the emitter circuit. From the equivalent generator theorem and Eq.~\r{inter} we obtain
\e
Z_m=-{{\cal E}_{12}\over I_1}={l_1^2l_2^2A^2_{ee}\over \omega^2Z_2}.
\l{Zm}\f

The final equivalent circuit of the radiating system where the presence of the nanoantenna is fully described by the mutual impedance $Z_m$ is depicted in Fig.~\ref{fig2}(b).
The radiation of the whole system is created by the current generator $I_1=I_e=j\omega d/l_1$ loaded by the series connection of the proper impedance $R_{\rm rad}+jX$ of the emitter and the mutual impedance $Z_m$. In the reactance $X$ of the emitter its proper $L$- and $C$-parameters are connected in series. In the mutual impedance $Z_m$ the effective mutual inductance $L_m$ and capacitance $C_m$ are connected in parallel. This difference needs to be explained.

The input impedance of the nanoantenna is a series connection of resistance $R_2$, inductance $L_2$, and capacitance $C_2$. Values $R_2,\ L_2$ and $C_2$ can be found from the Lorentzian model of the nanoantenna -- formulas \r{super}. Substituting $Z_2=R_2+j\omega L_2+1/j\omega C_2$, denoting $\omega_0=1/\sqrt{C_2L_2}$, and assuming that $\omega\approx \omega_0$, we may rewrite formula \r{Zm} as
\e
Z_m\approx {j\omega L_{\rm eff}\over 1-\left({\omega\over \omega_0}\right)^2-j\omega R_2C_2} N^2.
\l{parallel}\f
It is the standard formula of the circuit theory describing the impedance of a voltage transformer loaded by a low-loss parallel circuit resonating at $\omega_0$. In this formula $L_{\rm eff}=C_2\mu_0/\varepsilon_0$ is the effective inductance of the parallel circuit and the dimensionless value $N=\sqrt{\varepsilon_0} l_1l_2A_{ee}/\omega\sqrt{\mu_0}$ is an effective transformer parameter (called turns' ratio in the electrical engineering). In the vicinity of the resonance the dispersion of $Z_m$ is mainly determined by the denominator and we may neglect the frequency dependence
of the effective transformer putting in Eq.~\r{parallel} $N\approx \sqrt{\varepsilon_0} l_1l_2A_{ee}/\omega_0\sqrt{\mu_0}$. Then formula \r{parallel} describes the impedance of a parallel circuit with mutual inductance $L_{m}=\mu_0C_2N^2/\varepsilon_0$ and mutual capacitance $C_{m}=\varepsilon_0 L_2/N^2\mu_0$ connected to effective resistors (nonzero ones in both capacitive and inductive branches) which are responsible for the mutual resistance $R_m$.

The value $R_m\equiv R_{12}$ -- the real part of the right-hand side of Eq.~\r{parallel} comprises both radiative and dissipative resistance added to that of the emitter due to the presence of a nanoantenna. In the quasi-static approximation the value of $A_{ee}$ is real. Then $N$ is real and positive that results in the Purcell effect larger than unity. If $N\gg 1$ the Purcell factor at the resonance frequency may take huge values.

Since the driving current is fixed, the power delivered by the emitter to its environment is equal to $P=|I_1|^2 (R_{\rm rad}+R_m)$. The Purcell factor in accordance to \r{new1} takes the form
\e
F=1+{{\Re Z}_m\over R_{\rm rad}}=1+{6\pi l_1^2l_2^2\over \eta \omega^2 k^2l_1^2} \Re \left({A^2_{ee}\over Z_m}\right),
\l{Pur1}\f
where we have used formula \r{R_rad} for $R_{\rm rad}$ and substituted relation \r{Zm}. Now, applying the model of a Lorentzian scatterer to the nanoantenna we may express the impedance $Z_2$ of the nanoantenna through its polarizability
$\alpha_2\equiv\alpha_{NA}$. Really, $d_2={E_{21}l_2^2/ j\omega Z_2}$ and $\alpha_2=d_2/E_{21}$. Therefore, \r{Pur1} can be rewritten as
\e
F=1+{6\pi c^2\over \omega^3\eta \varepsilon_h}\Re\left(j\alpha_2 A^2_{ee}\right).
\l{Pur2}\f
This expression clearly shows that the Purcell factor does not depend on the emitter 1 -- only on the nanoantenna 2 and their mutual location.
Therefore we speak on the Purcell factor {\it of an object} at a point with radius vector $\-r_1-\-r_2$ with respect to the object. This factor is applicable to an arbitrary dipole emitter located at this point.

Now, it is time to make some other important comments. First, the problem of mutual coupling which we have solved above corresponds to the steady regime and is self-consistent at every frequency. The Purcell factor has the physical meaning at frequencies close to $\omega_0$, since the emission has a finite decay rate and its spectrum has nonzero bandwidth. Second, our analysis keeps valid in the case when the nanoantenna 2 has the resonance frequency $\omega_{02}$ different from the emission one $\omega_{0}\equiv \omega_{01}$. Still formula \r{Pur2} holds and the equivalent scheme depicted in Fig.~\ref{fig2} remains adequate, but the mutual impedance $Z_m$ is not anymore that of a simple parallel circuit connected through the transformer. However, if the difference between $\omega_0$ and $\omega_{20}$ is large, $\Im {\alpha}_2$ becomes too small
at the emission frequency $\omega_0$, and the Purcell factor is close to unity.

\begin{figure*}[!t]
\includegraphics[width=1.7\columnwidth]{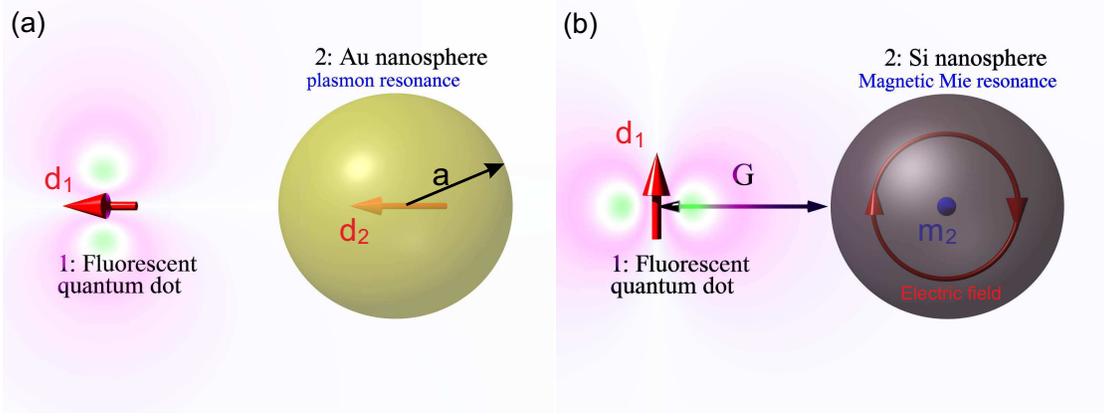}
\caption{(a) A fluorescent emitter over a plasmonic (e.g. golden) nanosphere has $F>1$ when its dipole moment is radially directed since in this case $A_{ee}>0$. (b) The same emitter over a dielectric (e.g. silicon) nanosphere has $F>1$ when its dipole moment is azimuthal: in this case $A_{em}>0$.}
\label{fig3}
\end{figure*}

The second comment is more important. In fact, the factor $A_{ee}$ (electric field of a unit dipole with origin $\-r_1$ evaluated at $\-r_2$) is complex due to the retardation effect. Its imaginary part is relevant for calculation of the Purcell factor at the frequencies different from the resonance frequency of object 2. Moreover, it is not exactly determined by the field of a unit dipole at the geometric center of the nanoantenna $\-r_{2g}$. The electromotive force induced by a point emitter in nanoantenna 2 may be found accurately -- via the integration of the local field $E_{21}({\-r})$ over the volume of the nanoantenna. If the local field is strongly non-symmetric with respect to its geometric center, the effective center ${\-r}_2$ of the nanoantenna shifts from the point $\-r_{2g}$ towards the emitter. Further, from the classical antenna theory~\cite{Balanis} it is known that for a two-element array of dipole antennas the mutual resistance is positive only when the antennas are collinear. This mutual location of dipoles 1 and 2 corresponds to Fig.~\ref{fig3}(a) when the dipole moment of the emitter is stretched radially towards a plasmonic nanosphere. In this case for small distances $G$ between the emitter and the sphere we may approximate $\Im A_{ee}=0$ and $A_{ee}\approx 1/2\pi\varepsilon_0\varepsilon_h D^3$, where $D=a+G$. In accordance to Eq.~\r{Pur2} this results in the Purcell factor higher than unity. However, if the dipole is located with respect to the nanosphere so that their dipole moments are parallel and not shifted, the interaction of dipoles becomes destructive. In this case one cannot neglect $\Im A_{ee}$, moreover, its contribution for distances $D$ comparable with $l_1$ and $l_2$ significantly exceeds that of the real part~\cite{Balanis}. Then the second term in \r{Pur2} becomes negative and makes the Purcell factor smaller than unity. This corresponds to the known situation: the mutual resistance $R_m$ of a transmitting dipole and a closely located reflector antenna is negative, and the enhancement of the directionality is accompanied by the decrease of the efficiency~\cite{Balanis}. For this case the antenna theory gives $R_{\rm rad}<|R_m|$~\cite{Balanis}. The energy balance is respected and we have $F>0$.

Last comment of this subsection refers to the approximation of the fixed dipole moment $d_1=d$. This approximation restricts the allowed strength of the dipole-dipole interaction by relatively modest values of the Purcell factor. Very high values of this factor would imply a very strong interaction of quantum object 1 (with both electromagnetic field and with nanoantenna 2). For this very strong coupling the approximation of predefined eigenstates becomes inadequate. Then, one has to solve the self-consistent problem for eigenmodes of the radiating system comprising the emitter and the nanoantenna. The solution within the framework of the semi-classical theory results in the so-called Rabi splitting~\cite{Rabi}. Instead of one emission frequency $\omega_0$ two emission frequencies arise, corresponding to two (one in-phase and one out-of-phase) spontaneous oscillations in the resonating and radiating system formed by objects 1 and 2. This effect has a well-known analogue in the classical theory of two coupled oscillators and can be of course described by a corresponding equivalent circuit. However, this phenomenon is beyond the framework of the present paper. Similarly, we do not consider the case of strong coupling when $\omega_{02}\ne \omega_{01}$.

\subsection{Validation of the Equivalent Circuit}

To validate our circuit model we apply it to an explicit structure depicted in Fig.~\ref{fig3}(a). First, let us show that formula \r{Pur2} based on the equivalent circuit fits the known analytical solution~\cite{Chew}. In that paper the Purcell factor was calculated using the exact solution of the electrodynamic problem of a dipole radiating in the presence of a sphere
of arbitrary radius $a$ filled by an isotropic material of (generally complex) permittivity $\varepsilon_s$. Formula (6) of that paper refers to the radial polarization of the dipole and its location outside the sphere. It is a series in which the first term corresponds to the dipole polarization of the sphere, i.e. for optically small spheres we may neglect the other terms.

In the framework of this approximation formula for the \textit{radiative} Purcell factor~\cite{Chew} reads:
\e
F\approx 9|j_1(kD)+b_1h_1^{(2)}(kD)|^2/(kD)^2,
\l{andrey}\f
where $j_1(X)$ and $h_1^{(2)}(X)$ are, respectively, spherical Bessel's and Hankel's functions with $n=1$, $D=a+G$ is the distance between the emitter and sphere centers, $k=q\sqrt{\varepsilon_h}$ is the wave number of the host medium and coefficient $b_1$ is given by formula~\cite{Chew}:
\e
b_1={\varepsilon_hj_1(ka)[k_saj_1(k_sa)]'-\varepsilon_sj_1(k_sa)[kaj_1(ka)]'
\over
\varepsilon_sj_1(k_sa)[kah_1^{(2)}(ka)]'-{\varepsilon_hh_1^{(2)}(ka)[k_saj_1(k_sa)]'}}.
\l{andrey1}\f
Here $k_s=q\sqrt{\varepsilon_s}$ is the wave number inside the sphere. The dipole approximation is valid when $|k_s|a\ll \pi$, practically when $|k_s|a<1$. To compare the radiative Purcell factor \r{andrey} with our result \r{Pur2} we have to remove losses that automatically equates the total Purcell factor to the radiative one. Therefore, we assume that $\varepsilon_s$ is real. Then $k_s$ is either real (if $\varepsilon_s>0$) or imaginary (if $\varepsilon_s<0$). In both these cases the following approximations are suitable for the spherical functions entering \r{andrey1}:
\e
j_1(X)\approx {X\over 3},\qquad h_1^{(2)}(X)\approx -{j\over X^2}, \quad X=k_sa,\, X=ka.
\l{approx}\f
With these substitutions the differentiation in \r{andrey1} becomes elementary, and we obtain for $b_1$ the result $b_1=jB$, where $B$ is a real value:
\e
B\approx {2 (ka)^3\over 3}{\varepsilon_s-\varepsilon_h\over \varepsilon_s+2\varepsilon_h}.
\l{andrey2}\f
Let us restrict the analysis by the case $kD\ll \pi$. Then, the asymptotic relations \r{approx} are suitable for $X=kD$. Formula \r{andrey} with substituted expressions \r{approx} and \r{andrey2} can be rewritten in a form
 \e
F\approx |1+j{3b_1^2\over (kD)^3}|^2=1+{9B^2\over q^6D^6\varepsilon_h^3}.
\l{andrey3}\f
Substitution of \r{andrey2} into \r{andrey3} results in
 \e
F\approx 1+{4a^6\over D^6}\left({\varepsilon_s-\varepsilon_h\over \varepsilon_s+2\varepsilon_h}\right)^2.
\l{andrey4}\f

Our circuit model resulted in formula \r{Pur2} which can be rewritten as
\e
F=1-{6\pi c^3\over \omega^3\sqrt{\varepsilon_h}}\Im\left(\alpha_2 A^2_{ee}\right).
\l{Pur3}\f
The quasi-static approximation for $A_{ee}$ has been already introduced:
\e
A_{ee}\approx {1\over 2\pi\varepsilon_0\varepsilon_h D^3}.
\l{aee}\f
Since the sphere is lossless, we have in accordance to \r{Lor}:
\e
\Im \alpha_2=-|\alpha_2|^2\Im \left({1\over \alpha_2}\right)=-\alpha_2^2 \left({k^3\over 6\pi\varepsilon_0\varepsilon_h}\right).
\l{Pur4}\f
The quasi-static polarizability $\alpha_{\rm QS}$ of a small sphere is well-known (see e.g. in~\cite{DellaSala2013}), and we have:
\e
\alpha_2\approx \alpha_{\rm QS} = 4\pi a^3\varepsilon_0\varepsilon_h{\varepsilon_s-\varepsilon_h\over \varepsilon_s+2\varepsilon_h}.
\l{Pur5}\f
Substituting \r{aee}, \r{Pur4} and \r{Pur5} into \r{Pur3} we obtain formula \r{andrey4}. Thus, the strict electrodynamic model and the present circuit model meet one another within the framework of the dipole approximation.

\begin{figure}[!b]
\includegraphics[width=1\linewidth]{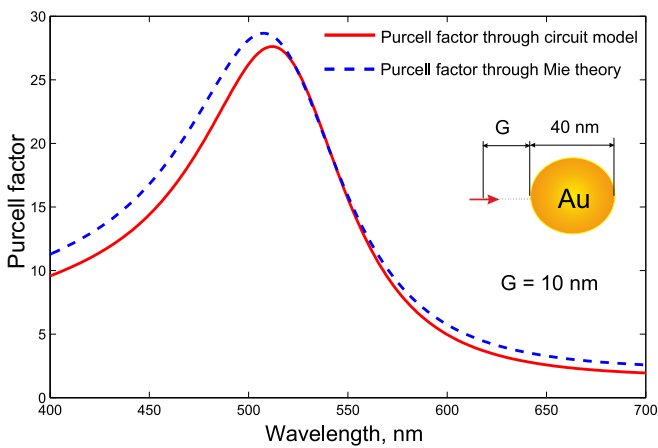}
\caption{The radial Purcell factor of a golden nanosphere of diameter $40$ nm at the distance $G=10$ nm in air: our circuit model (red solid curve) and exact Mie theory (blue dashed curve). Values of the permittivity are taken from the experiments of Johnson and
Christy~\cite{JC_72, MeyerDisp2006}.}
\label{fig4}
\end{figure}

In order to validate our circuit model for a more interesting case when the sphere is resonant (plasmonic nanoantenna) we consider an explicit structure of a golden sphere of diameter $2a=40$ nm. The radially polarized emitter is located at the distance $G=10$ nm from its surface. Values of the permittivity are taken from the experiments of Johnson and Christy~\cite{JC_72, MeyerDisp2006}. The radiating system is located in the air $\varepsilon_h=1$. In Fig.~\ref{fig4} we present our calculation of the Purcell factor performed with the use of \r{Pur3} in comparison with the Mie theory. In our calculation we have complemented formula \r{Pur5} by radiation losses in accordance to \r{alpha_inv_k3}:
$$
\alpha_2=\left({1\over \alpha_{\rm QS}}+j{k^3\over 6\pi\varepsilon_0\varepsilon_h}\right)^{-1}.
$$
For the plasmon resonance band our model is in agreement with the exact calculation. Our rough approximation for $A_{ee}$ works in this band because at the resonance the dipole eigenmode is realized. The excitation mechanism is not very important, and the sphere is polarized by an emitter as if it were excited by a plane wave -- nearly uniformly. The model becomes less accurate beyond the resonant band, where strong non-uniformity of the external field $E_{21}$ implies strong non-uniformity of the polarization decaying versus the distance from the emitter. Due to this decay the origin $\-r_2$ of the dipole $\-d_2$ shifts towards the emitter, the effective distance decreases compared to $D$ and $A_{ee}$ increases compared to \r{aee}. Therefore, it is not surprising that our model utilizing the simple approximation \r{aee} underestimates the Purcell effect at low frequencies.

\subsection{Extension of the Circuit Model}

Next, let us extend the equivalent scheme and generalize formula \r{Pur2}. First, let us see that the equivalent circuit keeps valid if the nanoantenna is a magnetic dipole. Of course, we mean artificial magnetism when the vortex polarization currents in the subwavelength particle result in its magnetic dipole moment. Qualitatively, this insight is applicable, for example, to a submicron silicon sphere at its magnetic Mie resonance (see e.g. in~\cite{Aizpurua_OE_2012, KrasnokOE, KrasnokNanoscale}). In Fig.~\ref{fig3}(b) we have depicted the corresponding radiating system. The $z$-directed electric dipole $d_1$ of the emitter 1 induces at the center of the nanosphere 2 a magnetic dipole $\-m_2=\-x_0m_2$ which is related to the local magnetic field via the magnetic polarizability $\beta_2\equiv m_2/H_x$. In this definition $H_x$ is the local field acting on the magnetic dipole. Here $H_x\equiv H_{21}$ is the magnetic field produced by the electric dipole $\-d_1$ in the plane $z=0$ at the distance $D$. It can be written as $H_{21}=j\omega A_{em} d_1$, where in the quasi-static limit $A_{em}\approx 1/4\pi D^2$.

The magnetic dipole antenna can be modeled as an optically small loop with an effective area $S$ and effective electric loop current $I$ which is considered uniform around the loop. The magnetic dipole moment is equal to $\mathbf {m}=\mu_0SI\mathbf{n}$, where $\mathbf{n}$ is a unit vector to the loop plane. The input impedance of the effective loop antenna equals to the ratio of IEMF ${\cal E}=j\omega \mu_0H_nS$ (where $H_n$ is the normal component of the local magnetic field) to the electric loop current $I$. So, in the present case the IEMF for the magnetic nanoantenna 2 resulting in the magnetic moment $\-m_2$ is equal to
${\cal E}_{21} =j\omega \mu_0 H_{21} S$, where $S$ is the effective area of the polarization current loop of the nanosphere (it will cancel out in the result). The induced magnetic moment $m_2=\mu_0{\cal E}_{12}S/Z_2$ comprises the factor $\omega^2$ in the magnetic analogue of \r{alpha_inv}:
\e {1\over \beta_2}={1\over \omega^2 S}\left(j\omega R_{2} -\omega X_2\right). \l{beta_inv}\f
Respectively, the Lorentzian model of the magnetic polarizability of a scatterer differs from the model of the electric polarizability by the factor $\omega^2$ (see e.g. in~\cite{Aizpurua_OE_2012}). The analogue of expression \r{Lor} takes form:
\e {1\over \beta}={1\over \omega^2\beta_0}(\omega_0^2-\omega^2+j\omega\Gamma_{\rm dis})+j{k^3\over 6\pi\mu_0}.\l{Lor1}\f
All other formulas of the Lorentzian model keep valid.
\begin{figure*}
\centering \epsfig{file=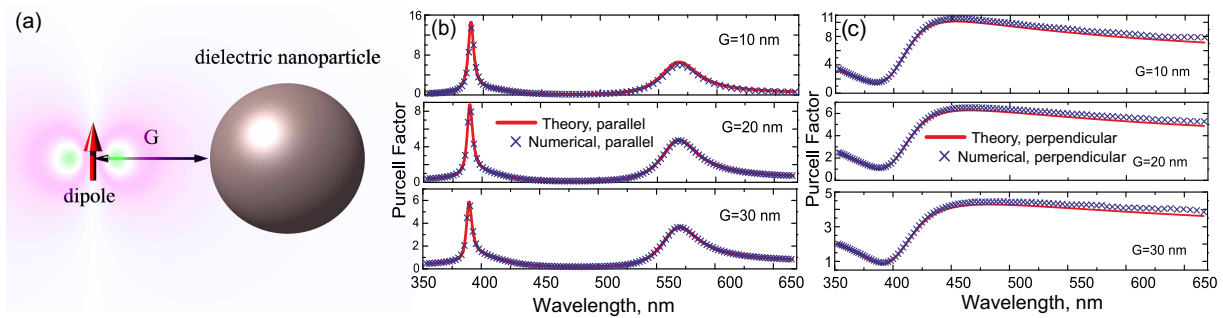,width=0.9\textwidth}
\caption{The Purcell factor extraction through a change of the input impedance in optics. (a) Illustration of a point dipole source located close to the dielectric spherical nanoparticle of the radius $a=70$~nm. (b) and (c) Purcell factor dependence on the emission wavelength for the parallel (b) and perpendicular (c) dipole orientation with respect to the sphere.} \label{opticalPurcell}
\end{figure*}

Accordingly, the equivalent circuit remains applicable. Magnetic moment $m_2$ produces the electric field $E_{12}=j\omega A_{em} m_2$, comprising due to the reciprocity the same coefficient $A_{em}$ which enters $H_{21}$. The corresponding IEMF ${\cal E}_{12}=E_{12}l_1$ is recalculated into the mutual impedance in the same way as above. Reproducing the same steps as for the electric dipole nanoantenna we come to the parallel-circuit formula \r{parallel} for $Z_m$ with substitution $N=\omega S_2l_1A_{em}/c$ for the transformer parameter. For the Purcell factor we obtain an analogue of \r{Pur2} in the form:
\e
F=1+{6\pi c^2\over \omega\eta }\Re\left(j\beta_2 A^2_{em}\right).
\l{Pur22}\f
If the response of the nanoantenna comprises both electric and magnetic dipoles, each of these modes is described by its own equivalent scheme. If these dipole moments resonate at the same frequency both equivalent circuits are similar and can be unified. A more complicated equivalent scheme would correspond to different resonances of the electric and magnetic modes. However, it is important to notice that both the electric and magnetic modes obviously contribute into the total mutual impedance, and both these contributions can be constructive. So, the excitation of an additional mode in the nanoantenna may increase $R_{\rm rad}$ enhancing $F$. The same refers to higher multipoles of the nanoantenna: each of the multipole modes contributes into total $Z_m$, and the coinciding or closely located resonances of high-order multipoles may result in huge values of the Purcell factor.

\subsection{Theoretical Verification of the General Approach}
\label{MethodOptics}

First, let us notice that our general approach resulted in formula \r{new1} is a useful alternative to the conventional methods of calculating the Purcell factor.
Although various numerical methods to solve the problems of nanophotonics and metamaterials have become widespread~\cite{MethodsBook}, direct numerical calculation of the Purcell factor using Green's function technique (\ref{PurcellImped}) or scattered field technique (\ref{PurcellQ}) faces fundamental difficulties. Indeed, the exact calculation of the microscopic field inside the quantum emitter as well as the exact calculation of the Green function at this point is challenging and time-consuming. Next, as shown in Ref.~\cite{Koenderink_2010} another known method of the Purcell factor calculation through the volume and quality factor of the cavity mode (see e.g. in works~\cite{Vahala_2003, Polman_PRB_2010, Cano_2013}) gives a strong disagreement with the accurate theoretical model (\ref{PurcellQ}), especially for plasmonic nanostructures. In finite systems and systems without losses the method of integrating the radiated power flow through some spherical surface surrounding the radiating system has become popular. However, in structures with losses this method depends on the choice of the integrating sphere (even low losses may strongly deviate the result since the integration surface is very large). Finally, all these methods can not be realized experimentally and extended to the radio frequency range (which is one of the purposes of the present study).

In commercial software packages, such as CST Studio~\cite{CST}, a point dipole can be modeled as an optically very short dipole of a perfectly conducting wire excited by an ideal current source. Since it has a finite length $l_1$, this dipole 1 in free space has a certain finite impedance, whose real part $R_{0,\rm in}$ is its radiation resistance. In presence of an arbitrary object 2 the IEMF ${\cal E}_{12}$ arises in the dipole and its input resistance modifies $R_{\rm in}\ne R_{0,\rm in}$. The input impedance results from exact simulations with the use of any reliable commercial software. The result for the Purcell factor $F=R_{\rm in}/R_{0,\rm in}$ should not depend on the length $l_1$ of the equivalent Hertzian dipole.
This method appears to be very practical and convenient for nanooptics. Moreover, it is more universal than all the aforementioned methods, because it is equally applicable to systems with or without losses.

To validate the general formula \r{new1} we have studied the structure depicted in Fig.~\ref{fig3}(b). In Fig.~\ref{opticalPurcell}(a) the geometry of the problem under consideration is recalled. The quantum source is modeled in CST Microwave Studio as a Hertzian dipole of length 10 nm. The dielectric spherical nanoparticle of radius $a=$70 nm and relative permittivity 15 is located at distance $G$ from the dipole. The Purcell factor retrieved from numerical simulations as $F=R_{\rm in}/R_{0,\rm in}$ is compared with the exact solution~\cite{Chew} in which now the series has been accurately evaluated.
We studied both parallel and orthogonal dipole orientations, corresponding to Figs.~\ref{opticalPurcell}(b) and (c), respectively, for three values of $G$. The exact solution and our results are in excellent agreement. For the orthogonal orientation at wavelength $\lambda\approx 570$ nm the sphere experiences the magnetic Mie resonance, and at $\lambda\approx 390$ nm -- the electric dipole and magnetic quadrupole (makes the largest contribution) Mie resonances. Unfortunately, our simplistic model resulting in formulas \r{Pur2} and \r{Pur22} does not offer enough numerical accuracy due to two factors. First, the electric dipole mode cannot be neglected at the magnetic resonance. Second, the electric resonance holds at higher frequency, where the electromagnetic response of the nanosphere is not purely dipolar. However, for our current purpose it is enough that the exact version of our method -- formula~\r{new1} -- gives an excellent accuracy.

\subsection{Purcell Factor for Radio Antennas}

\begin{figure}[!t]
\centering \epsfig{file=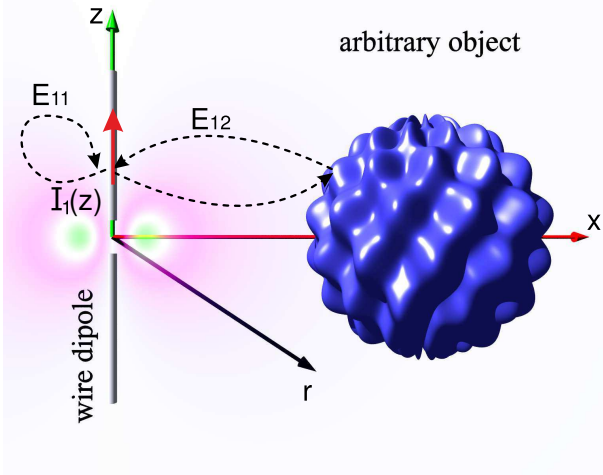,width=0.47\textwidth}
\caption{A schematic illustration of the radiative dipole antenna placed near an arbitrary scattering object. Each current density element of the antenna interacts with itself and other elements of the current ($\mathbf{E}_{11}$), as well as with an object ($\mathbf{E}_{12}$).} \label{geom}
\end{figure}

Now, let us go beyond the optical frequency range and extend the whole concept to radio frequencies, including microwaves, millimeter waves and teraherz frequency range. Instead of a quantum emitter let us consider a dipole antenna 1 interacting with an arbitrary object 2, as it is sketched in Fig.~\ref{geom}. If the dipole 1 is resonant, e.g. has length $l_1=\lambda/2$ it can be excited by a short pulse (an analogue of the optical pumping) and will irradiate its energy at its resonant frequency $\omega_0$ during the finite emission time $1/\gamma_0$. If the radiation quality of the antenna is high the time $1/\gamma_0$ is very long in terms of the period $2\pi/\omega_0$. It may be reduced to $1/\gamma\ll 1/\gamma_0$ if an object 2 is located in the vicinity of the antenna 1 which increases its radiation resistance. Object 2 is not obviously a resonator tuned to the same frequency as it is adopted in optical applications of the Purcell effect. In accordance to
our consideration in subsection \label{GeneralMethod} it can be an arbitrary object constructively interacting with the antenna. Then the general equivalent circuit shown in Fig.~\ref{fig1}(b) remains valid and results in the mutual impedance $Z_m$. Of course, in the general case the mutual impedance is not obviously that of a parallel $RLC$-circuit. What is essential that $\Re Z_m\equiv R_{12}$ should be positive and increase the input resistance of antenna 1. The impressed current $I_e\equiv I_1$ is then determined by the dipole moment of the antenna 1 at the moment when the external pulse ends. In fact, this consideration, and the representation of the object 2 via the mutual impedance $Z_m$ have been known in antenna engineering for a long time, see e.g.~\cite{VanBladel}.

It is difficult to significantly increase the radiation resistance of an already efficient antenna -- that with the resonant length $l_1=\lambda/2$. Absolute values of $R_m$ may be noticeable in this case, but the relative contribution will be modest. The concept of the Purcell factor becomes relevant for a short dipole -- that with a low radiation resistance $R_{0,\rm rad}$, much lower than the internal resistance of the voltage generator applied to the radio antenna. As a rule, this is the output resistance of the feeding transmission line which usually equals $R_{\rm out}=50$ Ohms. If $R_{0,\rm rad}\ll R_{\rm out}$, the presence of a low-loss object inserting positive mutual resistance may lead to much better matching of the effective generator to the antenna and therefore to much higher radiation. At first glance, this radiation gain has nothing to do with the Purcell factor, which describes the emission regime. However, for a very short dipole $l_1\ll \lambda/2$
these values are equal to one another.

The proper reactance of a short dipole antenna is capacitive. The spontaneous emission (quasi-harmonic radiation after a short pulse) is possible if the output impedance of the feeding line has the inductive reactance connected in series with $R_{\rm out}$. Then, in the absence of object 2 the emission is still described by the current source $I_1$, at the frequency $\omega_0=1/\sqrt{LC}$ loaded by the resistance of the feeding line $R_{\rm out}=50$ Ohms and the radiation resistance $R_{0,\rm rad}\ll R_{\rm out}$. Most part of the energy is lost in $R_{\rm out}$ and only a small portion of the pulse energy is irradiated. The presence of object 2 changes this distribution increasing the radiation resistance and the decay rate multiplies by $F=R_{\rm rad}/R_{0,\rm rad}$.

In the regime of the usual transmission at the frequency $\omega_0$, the steady-state voltage $V$ at the output of the feeding line is loaded by the resistance $R_{\rm out}=50$~Ohms and the antenna input impedance $Z_{\rm in}$. The input impedance of a small antenna consists of a small radiation resistance $R_{0,\rm rad}\ll R_{\rm out}$ and a very high reactance $X$. In this case the current $I_1=V/(R_{\rm out}+R_{0,\rm rad}+jX)\approx V/(R_{\rm out}+jX)\ll V/R_{0,\rm rad}$ and only a small portion of the supplied power is radiated. The power is mainly reflected from the antenna back to the generator. The presence of object 2 increases $R_{\rm rad}$, i.e. improves the matching of the antenna to the feeding line. The radiated power increases in accordance to formula $P_{\rm rad}=|I_1|^2R_{\rm rad}$. However, matching remains poor since $R_{\rm rad}\ll |R_{\rm out}+jX|$. Therefore, we can write $I_1=V/(R_{\rm out}+R_{\rm rad}+jX)\approx V/(R_{\rm out}+jX)$. The radiating current does not change in the presence of object 2 though the input resistance of the antenna 1 changes! The increase of the radiated power is solely described by the increase of the radiation resistance. Therefore, the gain in the transmitted radiation is equal to the Purcell factor $F=R_{\rm rad}/R_{0,\rm rad}$.

Briefly, for a very poor transmitting antenna 1 we may find the Purcell factor of object 2
from the usual radiation gain of the same antenna 1 in presence of the object 2. This factor describes the emission of the pulse energy by antenna 1 in presence of the radiation-enhancing object. It does not depend on the antenna itself and is fully determined by the properties of the object and its location.
Vice versa, we can predict how much the antenna will radiate due to the presence of the object if this antenna is tuned into resonance, excited in the absence and presence of the object by a pulse voltage, and find the decay rate of its emission after the pulse is gone.
We should stress that the Purcell factor of object 2 measured with the use of an antenna is not the same as the radiation enhancement of this antenna in the presence of object 2. Only in an important special case when the probe antenna is a very poor emitter they are approximately equal. The observation of this equivalence dramatically extends the field where the notion of the Purcell factor is relevant.

Our last extension concerns the Purcell factor of an arbitrary object acting on a magnetic dipole antenna. We have already noticed that the magnetic dipole antenna is an optically small loop (can be multi-turn~\cite{Balanis}) with an effective area $S$ and electric current $I$ which is practically uniform around the loop. The magnetic dipole moment
$\mathbf {m}=\mu_0SI\mathbf{n}$, where $\mathbf{n}$ is a unit vector to the loop plane is related to the effective magnetic current as $I_m={\dot{m}}=-j\omega{m}$. The input impedance of the loop antenna equals to the ratio of IEMF ${\cal E}=j\omega \mu_0H_nS$ (where $H_n$ is the normal component of the local magnetic field) to the induced electric current $I$ and can be rewritten as $Z_{\rm in,m}=H_{n}/I_{\rm m}$. This offers a full analogy with the electric dipole antenna and corresponds to the duality principle. It is clear that the input impedance of the magnetic antenna is related to the Green function at the magnetic dipole origin as $G_{zz}(0,0,\omega)=-j\omega Z_{\rm in,m}/q^{2}$. After extracting the imaginary part from the last expression, we obtain the Purcell factor in the form \r{new1} ~\cite{Slobozhanyuk2014magnetic}. So, all the theory developed above including the equivalent circuits keeps valid.

\section{Measurement of the Purcell factor in the Microwave Spectral Range}
\label{Methodexper}

\begin{figure*}[!t]
\includegraphics[width=2.0\columnwidth]{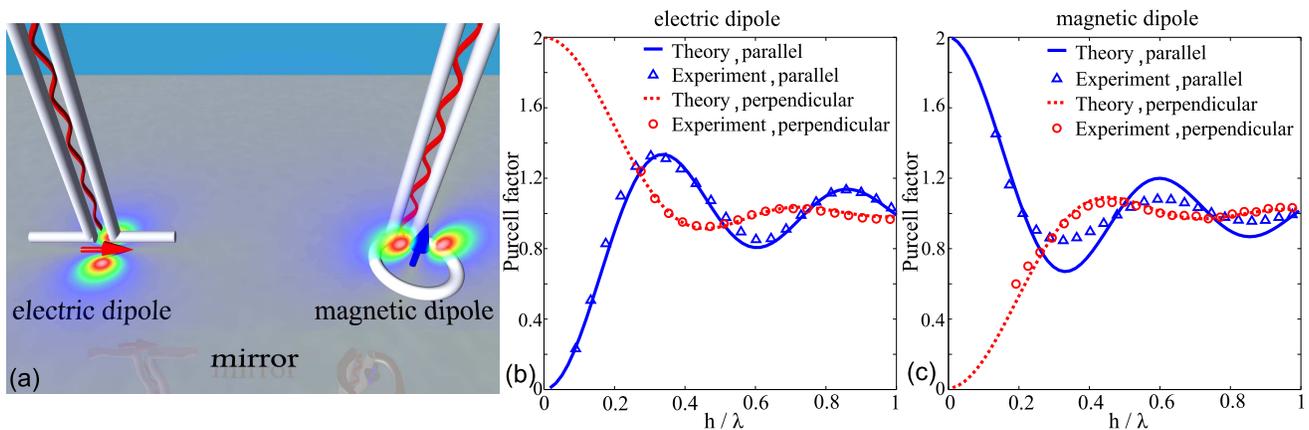}
\caption{(a) Geometry of the experiment to measure the value of the Purcell factor for an electric dipole antenna near a perfect metallic mirror. (b) Measured results for the Purcell factor (symbols) along with the analytical results Eq.~(\ref{PurcellElMag}) for parallel and perpendicular orientations of the electric dipole antenna with respect to the mirror. (c) The same results for the magnetic dipole antenna.}
\label{GeomFig1}
\end{figure*}

Now let us demonstrate the application of our method experimentally, retrieving the Purcell factor from measured input resistance of a radio antenna using Eq.~\r{new1}.
The input impedance of an antenna can be easily determined from the $S$-parameters. Namely, for a dipole antenna connected to a one-mode waveguide (e.g. a coaxial cable), the quantity $R_{\rm in}$ is related to the reflection coefficient $S_{11}$ measured at the waveguide input and the characteristic impedance of the waveguide $Z_{\rm w}$~\cite{Balanis}:
\e \label{RIN} R_{\rm in}=Z_{\rm w}\frac{1-[\Re (S_{11})]^2-[\Im (S_{11})]^2}{[1-\Re (S_{11})]^2+[\Im (S_{11})]^2}\:.\f

In our experimental verification of the technique object 2 is a flat copper plate of optically large size and the antenna 1 is located near its center. This plate in the microwave range emulates the
perfectly conducting plane and the Purcell effect in this case may be referred to as a special case of spontaneous emission near an interface~\cite{Noginova2014, snoeks1995, blum2012, Narimanov_APL_2012, Drexhage_1970, Xiao_2011, Hughes_2009, Haglund_2001, Gresillon_2008, Sipe_1984, Abdi_2013, Kunz_JOSA_77}. For the perfectly conducting interface a simple analytical result for the Purcell factor was obtained in~\cite{Sipe_1984, Piche_1992}. The expression for the electric ($F_{\rm e}$) and magnetic ($F_{\rm m}$) Purcell factor for either parallel ($\|$) or perpendicular ($\bot$) orientations are as follows:
\begin{eqnarray}\label{PurcellElMag}
F_{\rm e,m}^{\bot} &=&
1\pm3\left[\frac{\sin(\eta)}{\eta^3}-\frac{\cos(\eta)}{\eta^2}\right]\:,\nonumber\\
F_{\rm e,m}^{\|} &=&1\mp\frac{3}{2}\left[\frac{\cos(\eta)}{\eta^2}+\left[\frac{1}{\eta}-\frac{1}{\eta^3}\right]\sin(\eta)\right]\:,
\end{eqnarray}
where $\eta=2qh$, $h$ is the height of the (electric or magnetic) dipole above the metal. The upper sign corresponds to an electric dipole, the lower to the magnetic one.
We have compared the predictions of Eqs.~\eqref{PurcellElMag} with $F=R_{\rm in}/R_{0,\rm in}$. The experimental setup is schematically shown in Fig.~\ref{GeomFig1}. As two stems of an electric dipole antenna we use brass wires of length 0.4~cm soldered to the internal and external veins of the coaxial cable connected to a vector network analyzer. The wave impedance of the cable is equal to $Z_{\rm w}=50$ $\Omega$ that guarantees the regime $R_{0,\rm rad}\ll Z_{\rm w}$. Magnetic dipole source is realized as a wire ring with the diameter 1 cm connected similarly. The measurement is performed in the spectral range 5--14 GHz which corresponds to wavelengths from 2.14 to 6 cm. The object 2 was a polished stainless steel sheet with sides $180\times 210$ cm (the smallest mirror side greatly exceeds the largest wavelength and the diffraction effects are negligible). The antennas were attached to an arm of a precise coordinate scanner which
moved in the vertical directions, allowing us to measure the Purcell factor as a function of the emitter height. The main experimental results for electric and magnetic antennas are shown in Fig.~\ref{GeomFig1}b,c (squares and triangles correspond to two orientations of the electric and magnetic antennas). The solid blue and dashed red curves represent the theoretical values of the Purcell factor (\ref{PurcellElMag}). Experimental and theoretical results are in excellent agreement. The Purcell factor exhibits oscillations with the period on the order of the wavelength when the source is moved vertically. These oscillations are due to the interference pattern which exhibits in the radiation resistance $R_{\rm in}\approx R_{\rm rad}$. It clearly indicates that our general formula \r{new1} is applicable far beyond the quasi-static interaction between objects 1 and 2 assumed in the previous section. When $h$ increases $F$ eventually saturates at unity.

Slight disagreement can be noticed for the magnetic antenna. It is explained by a slight current inhomogeneity around the ring. This inhomogeneity appears when the magnetic dipole is parallel to the metal plane i.e. the loop is in the vertical plane. Definitely the lower half of the loop is stronger capacitively coupled to the metal plane than the upper one, and it results in this inhomogeneity. Notice, that for very small $h$ we could not measure the Purcell factor due to the finite size of the our antennas when Eqs.~\eqref{PurcellElMag} become inapplicable. Interesting, that the deviations from these formulas were also observed for a quantum dot, located too closely to the mirror~\cite{Lodahl2011}.

Importantly, in the microwave frequency range the {electric} dipole antenna is usually fed by a coaxial cable whose thickness is not negligible, though optically very small. This factor results in the radiation from the cable open end and affects the measured Purcell factor. We directly measure not the input resistance $R_{\rm in}$ of the antenna but the sum of $R_{\rm in}$ and $\delta R$, where the last term in the radiation resistance of the open cable. Therefore, we have separately measured the input resistance of the open end of the cable which obviously equals to $\delta R$ and subtracted it from $R_{\rm in}$ found with the use of Eq.(\ref{RIN}). Otherwise, the disagreement in Fig.~\ref{GeomFig1}b,c would be more noticeable.

\section*{Conclusions}
In this paper, we have analyzed theoretically and experimentally the classical counterpart of the Purcell effect for subwavelength electric and magnetic dipole antennas. We have generalized the approach accepted in nanophotonics to the case of microwave antennas and recovered the known expression for the Purcell factor via the imaginary part of the electromagnetic Green's function. Using this result, we propose a new method to directly measure the Purcell factor through the input impedance of small antenna. We have experimentally verified the technique for both electric and magnetic dipole antennas. The technique has been also successfully applied to a cornerstone problem of the all-dielectric nanophotonics: Purcell effect due to the Mie resonances of dielectric sphere. We believe that the proposed Purcell factor extraction method is versatile and can be used in various frequency ranges: from radio to optics.

\section*{Acknowledgements}
The authors acknowledge useful discussions with I.S. Maksymov, I.V. Iorsh, S.B. Glybovski and P. Ginzburg. We are also grateful to D.S. Filonov for his interest in this work. This work was supported by the Ministry of Education and Science of the Russian Federation (projects 14.584.21.0009 10, GOSZADANIE 2014/190, Zadanie no. 3.561.2014/K), Russian Foundation for Basic Research, "Dynasty" Foundation (Russia), the Australian Research Council via Future Fellowship program (FT110100037), and the Australian National University.


\begin{thebibliography}{10}

\bibitem{Purcell_46}
E.~M. Purcell.
\newblock Spontaneous emission probabilities at radio frequencies.
\newblock {\em Phys. Rev.}, 69:681, 1946.

\bibitem{Maksymov_PRL_2013}
C.~Sauvan, J.~P. Hugonin, I.~S. Maksymov, and P.~Lalanne.
\newblock Theory of the spontaneous optical emission of nanosize photonic and
  plasmon resonators.
\newblock {\em Physical Review Letters}, 110:237401, 2013.

\bibitem{Moerner_bowtie_09}
Anika Kinkhabwala, Zongfu Yu, Shanhui Fan, Yuri Avlasevich, Klaus Mullen, and
  W.~E. Moerner.
\newblock Large single-molecule fluorescence enhancements produced by a bowtie
  nanoantenna.
\newblock {\em Nature Photonics}, 3:654--657, 2009.

\bibitem{Vahala_2003}
Kerry~J. Vahala.
\newblock Optical microcavities.
\newblock {\em Nature}, 424:839--846, 2003.

\bibitem{Asano_2007}
Susumu Noda, Masayuki Fujita, and Takashi Asano.
\newblock Spontaneous-emission control by photonic crystals and nanocavities
  (review).
\newblock {\em Nature Photonics}, 1:449 -- 458, 2007.

\bibitem{Hu_NAT_2012}
Kasey~J. Russell, Tsung-Li Liu, Shanying Cui, and Evelyn~L. Hu.
\newblock Large spontaneous emission enhancement in plasmonic nanocavities.
\newblock {\em Nature Photonics}, 6:459--462, 2012.

\bibitem{Waks_2013}
Chad Ropp, Zachary Cummins, Sanghee Nah, John~T. Fourkas, Benjamin Shapiro, and
  Edo Waks.
\newblock Nanoscale imaging and spontaneous emission control with a single
  nano-positioned quantum dot.
\newblock {\em Nature Communications}, 4:1447, 2013.

\bibitem{Carminati_PRL_14}
L.~Aigouy, A.~Caze, P.~Gredin, M.~Mortier, and R.~Carminati.
\newblock Mapping and quantifying electric and magnetic dipole luminescence at
  the nanoscale.
\newblock {\em Physical Review Letters}, 113:076101, 2014.

\bibitem{Iorsh_2013}
Alexander Poddubny, Ivan Iorsh, Pavel Belov, and Yuri Kivshar.
\newblock Hyperbolic metamaterials.
\newblock {\em Nature Photonics}, 7:948--957, 2013.

\bibitem{Cano_2013}
Mario Agio and Diego~Martin Cano.
\newblock Nano-optics: The purcell factor of nanoresonators.
\newblock {\em Nature Photonics}, 7:674--675, 2013.

\bibitem{kavbamalas}
A.~Kavokin, J.J. Baumberg, G.~Malpuech, and F.P. Laussy.
\newblock {\em {M}icrocavities}.
\newblock Clarendon Press, Oxford, 2006.

\bibitem{Fainman_2010}
Maziar~P. Nezhad, Aleksandar Simic, Olesya Bondarenko, Boris Slutsky, Amit
  Mizrahi, Liang Feng, Vitaliy Lomakin, and Yeshaiahu Fainman.
\newblock Room-temperature subwavelength metallo-dielectric lasers.
\newblock {\em Nature Photonics}, 4:395--399, 2010.

\bibitem{Fainman_2013}
Qing Gu, Boris Slutsky, Felipe Vallini, Joseph S.~T. Smalley, Maziar~P. Nezhad,
  Newton~C. Frateschi, and Yeshaiahu Fainman.
\newblock Purcell effect in sub-wavelength semiconductor lasers.
\newblock {\em Optics Express}, 21:15603--15617, 2013.

\bibitem{Sandoghdar_Nature_00}
J.~Michaelis, C.~Hettich, J.~Mlynek, and V.~Sandoghdar.
\newblock Optical microscopy using a single-molecule light source.
\newblock {\em Nature}, 405:325, 2000.

\bibitem{Koenderink_PRL_11}
Martin Frimmer, Yuntian Chen, and A.~Femius Koenderink.
\newblock Scanning emitter lifetime imaging microscopy for spontaneous emission
  control.
\newblock {\em Physical Review Letters}, 107:123602, 2011.

\bibitem{Cosa_2013}
Gonzalo Cosa.
\newblock Single-molecule fluorescence: Assembling nanoantennas.
\newblock {\em Nature Chemistry}, 5:159--160, 2013.

\bibitem{Wilde_2014}
D.~Cao, A.~Caze, M.~Calabrese, R.~Pierrat, N.~Bardou, S.~Collin, R.~Carminati,
  V.~Krachmalnicoff, and Y.~De Wilde.
\newblock Mapping the radiative and non-radiative local density of states in
  the near-field of a gold nanoantenna.
\newblock {\em arXiv:1401.2858v1}, 2014.

\bibitem{Vamivakas_NanoLetters_13}
Ryan Beams, Dallas Smith, Timothy~W. Johnson, Sang-Hyun Oh, Lukas Novotny, and
  A.~Nick Vamivakas.
\newblock Nanoscale fluorescence lifetime imaging of an optical antenna with a
  single diamond nv center.
\newblock {\em Nano Letters}, 13:3807--3811, 2013.

\bibitem{Tinnefeld_Science_2012}
G.~P. Acuna, F.~M. Moller, P.~Holzmeister, S.~Beater, B.~Lalkens, and
  P.~Tinnefeld.
\newblock Fluorescence enhancement at docking sites of dna-directed
  self-assembled nanoantennas.
\newblock {\em Science}, 338:506--510, 2012.

\bibitem{Kumar2013}
Nikhil Kumar.
\newblock {\em Spontaneous Emission Rate Enhancement Using Optical Antennas}.
\newblock PhD thesis, University of California at Berkeley, 2013.

\bibitem{Novotny_Hecht_book}
L.~Novotny and B.~Hecht.
\newblock {\em Principles of Nano-Optics}.
\newblock Cambridge University Press, 2006.

\bibitem{Khitrova2006}
G.~{Khitrova}, H.~M. {Gibbs}, M.~{Kira}, S.~W. {Koch}, and A.~{Scherer}.
\newblock {V}acuum {R}abi splitting in semiconductors.
\newblock {\em Nature Physics}, 2:81--90, February 2006.

\bibitem{Book_AtomMol}
W.~Demtr\"oder.
\newblock {\em Atoms, Molecules and Photons: An Introduction to Atomic-,
  Molecular- and Quantum Physics}.
\newblock Springer, 2006.

\bibitem{Lodahl2011}
M.~L. {Andersen}, S.~{Stobbe}, A.~S. {S{\o}rensen}, and P.~{Lodahl}.
\newblock {Strongly modified plasmon-matter interaction with mesoscopic quantum
  emitters}.
\newblock {\em Nature Physics}, 7:215--218, March 2011.

\bibitem{Kavokin1991}
E.~L. Ivchenko and A.~V. Kavokin.
\newblock {L}ight {R}eflection from {Q}uantum {W}ell, {Q}uantum {W}ire and
  {Q}uantum {D}ot {S}tructures.
\newblock {\em Sov.Phys.Solid State}, 34(6):1815--1822, 1992.

\bibitem{lagendijk_review}
Pedro {de Vries}, David~V. {van Coevorden}, and Ad~Lagendijk.
\newblock {P}oint scatterers for classical waves.
\newblock {\em Rev. Mod. Phys.}, 70(2):447--466, Apr 1998.

\bibitem{barnett1996}
Stephen~M Barnett, Bruno Huttner, Rodney Loudon, and Reza Matloob.
\newblock {D}ecay of excited atoms in absorbing dielectrics.
\newblock {\em J. Phys. B}, 29(16):3763, 1996.

\bibitem{Welsch2006}
W.~Vogel and D.-G. Welsch.
\newblock {\em {Q}uantum {O}ptics}.
\newblock Wiley, Weinheim, 2006.

\bibitem{Balanis}
C.~Balanis.
\newblock {\em Antenna theory: analysis and design}.
\newblock New York; Brisbane: J. Wiley, 1982.

\bibitem{new}
Jeppe Johansen, Soren Stobbe, Ivan~S. Nikolaev, Toke Lund-Hansen, Philip~T.
  Kristensen, Jorn~M. Hvam, Willem~L. Vos, and Peter Lodahl.
\newblock Size dependence of the wavefunction of self-assembled inas quantum
  dots from time-resolved optical measurements.
\newblock {\em Physical Review B}, 77:073303(1--6), 2008.

\bibitem{Boucaud_2010}
X.~Checoury, Z.~Han, M.~El Kurdi, and P.~Boucaud.
\newblock Deterministic measurement of the purcell factor in microcavities
  through raman emission.
\newblock {\em Physical Review A}, 81:033832, 2010.

\bibitem{NovotnyAntennasForLight}
Lukas Novotny and Niek van Hulst.
\newblock Antennas for light.
\newblock {\em Nature Photonics}, 5:83-90, 2011.

\bibitem{Krasnok_UFN_2013}
A.E. Krasnok, I.S. Maksymov, A.I. Denisyuk, P.A. Belov, A.E. Miroshnichenko,
  C.R. Simovski, and Yu.S. Kivshar.
\newblock Optical nanoantennas.
\newblock {\em Phys.-Usp.}, 56:539, 2013.

\bibitem{Raschke_2012}
R.~L. Olmon and M.~B. Raschke.
\newblock Antenna-load interactions at optical frequencies: impedance matching
  to quantum systems.
\newblock {\em Nanotechnology}, 23:444001, 2012.

\bibitem{Alu_dip_PRL_08}
Andrea Alu and Nader Engheta.
\newblock Input impedance, nanocircuit loading, and radiation tuning of optical
  nanoantennas.
\newblock {\em PRL}, 101:043901, 2008.

\bibitem{Hecht_009}
Jer-Shing Huang, Thorsten Feichtner, Paolo Biagioni, and Bert Hecht.
\newblock Impedance matching and emission properties of nanoantennas in an
  optical nanocircuit.
\newblock {\em Nano Letters}, 9:1897-1902, 2009.

\bibitem{Marquier_PRL_2010}
Jean-Jacques Greffet, Marine Laroche, and Francois Marquier.
\newblock Impedance of a nanoantenna and a single quantum emitter.
\newblock {\em Physical Review Letters}, 105:117701, 2010.

\bibitem{ScullyZubairy}
Marian~O. Scully and M.~Suhail Zubairy.
\newblock {\em {Q}uantum {O}ptics}.
\newblock Cambridge University Press, Cambridge, UK, 1997.

\bibitem{GinzburgUFN}
V.L. Ginzburg.
\newblock On the nature of spontaneous emission.
\newblock {\em Uspekhi Fizicheskikh Nauk}, 140:687-698, 1983.

\bibitem{Ghiner_2000}
G.~I. Surdutovicha and A.~V. Ghiner.
\newblock A two-level atom and the problem of the radiation reaction in the
  semiclassical theory: optical bloch equations revisited.
\newblock {\em Physica A: Statistical Mechanics and its Applications},
  283:212-217, 2000.

\bibitem{Slobozhanyuk2014magnetic}
A.~P. Slobozhanyuk, A.~N. Poddubny, A.~E. Krasnok, and P.~A. Belov.
\newblock Magnetic purcell factor in wire metamaterials.
\newblock {\em Applied Physics Letters}, 104:161105, 2014.

\bibitem{Krasnok_superdir_APL}
Alexander~E. Krasnok, Dmitry~S. Filonov, Constantin~R. Simovski, Yuri~S.
  Kivshar, and Pavel~A. Belov.
\newblock Experimental demonstration of superdirective dielectric antenna.
\newblock {\em Applied Physics Letters}, 104:133502, 2014.

\bibitem{Lenac}
M.~S. Tomas and Z.~Lenac.
\newblock Decay of excited molecules in absorbing planar cavities.
\newblock {\em Physical Review A}, 56:4197-4206, 1997.

\bibitem{Jackson}
J.D. Jackson.
\newblock {\em Classical Electrodynamics}.
\newblock New York : Wiley, 1998.

\bibitem{Tret}
S.~Tretyakov.
\newblock Maximizing absorption and scattering by dipole particles.
\newblock {\em Plasmonics}, 9:935--944, 2014.

\bibitem{Rabi}
G.~Khitrova, H.~M. Gibbs, M.~Kira, S.~W. Koch, and A.~Scherer.
\newblock Vacuum rabi splitting in semiconductors.
\newblock {\em Nature Physics}, 2:81--89, 2006.

\bibitem{Chew}
H.~Chew.
\newblock Transition rates of atoms near spherical surfaces.
\newblock {\em Journal of Chemical Physics}, 87:1355--1360, 1987.

\bibitem{DellaSala2013}
Fabio~Della Sala and Stefania D'Agostino, editors.
\newblock {\em Handbook of Molecular Plasmonics}.
\newblock CRC Press, 2013.

\bibitem{JC_72}
P.~B. Johnson and R.~W. Christy.
\newblock Optical constants of the noble metals.
\newblock {\em Physical Review B}, 6:4370, 1972.

\bibitem{MeyerDisp2006}
P.~G. Etchegoin, E.~C.~Le Ru, and M.~Meyer.
\newblock An analytic model for the optical properties of gold.
\newblock {\em J. Chem. Phys.}, 125:164705, 2006.

\bibitem{Aizpurua_OE_2012}
M.~K. Schmidt, R.~Esteban, J.~J. Saenz, I.~Suarez-Lacalle, S.~Mackowski, and
  J.~Aizpurua.
\newblock Dielectric antennas - a suitable platform for controlling magnetic
  dipolar emission.
\newblock {\em Optics Express}, 20:13636--13650, 2012.

\bibitem{KrasnokOE}
A.~E. Krasnok, A.~E. Miroshnichenko, P.~A. Belov, and Yu.~S. Kivshar.
\newblock All-dielectric optical nanoantennas.
\newblock {\em Optics Express}, 20:20599, 2012.

\bibitem{KrasnokNanoscale}
Alexander~E. Krasnok, Constantin~R. Simovski, Pavel~A. Belov, and Yuri~S.
  Kivshar.
\newblock Superdirective dielectric nanoantenna.
\newblock {\em Nanoscale}, 6:7354--7361, 2014.

\bibitem{MethodsBook}
Kenneth Diest, editor.
\newblock {\em Numerical Methods for Metamaterial Design}, volume 127 of {\em
  Topics in Applied Physics}.
\newblock Springer, 2013.

\bibitem{Koenderink_2010}
A.~F. Koenderink.
\newblock On the use of purcell factors for plasmon antennas.
\newblock {\em Optics Letters}, 35:4208--4210, 2010.

\bibitem{Polman_PRB_2010}
Ernst Jan~R. Vesseur, F.~Javier~Garcia de~Abajo, and Albert Polman.
\newblock Broadband purcell enhancement in plasmonic ring cavities.
\newblock {\em Physical Review B}, 82:165419, 2010.

\bibitem{CST}
{\em CST STUDIO SUITE, www.cst.com}, 2014.

\bibitem{VanBladel}
Jean G.~Van Bladel.
\newblock {\em Electromagnetic Fields, 2nd Edition}.
\newblock Wiley-IEEE Press, 2007.

\bibitem{Noginova2014}
R.~Hussain, D.~Keene, N.~Noginova, and M.~Durach.
\newblock Spontaneous emission of electric and magnetic dipoles in the vicinity
  of thin and thick metal.
\newblock {\em Optics Express}, 22(7):7744--7755, Apr 2014.

\bibitem{snoeks1995}
E.~Snoeks, A.~Lagendijk, and A.~Polman.
\newblock Measuring and modifying the spontaneous emission rate of erbium near
  an interface.
\newblock {\em Physical Review Letters}, 74:2459--2462, Mar 1995.

\bibitem{blum2012}
Christian Blum, Niels Zijlstra, Ad~Lagendijk, Martijn Wubs, Allard~P. Mosk,
  Vinod Subramaniam, and Willem~L. Vos.
\newblock Nanophotonic control of the f\"orster resonance energy transfer
  efficiency.
\newblock {\em Physical Review Letters}, 109:203601, Nov 2012.

\bibitem{Narimanov_APL_2012}
Zubin Jacob, Igor~I. Smolyaninov, and Evgenii~E. Narimanov.
\newblock Broadband purcell effect: Radiative decay engineering with
  metamaterials.
\newblock {\em Applied Physics Letters}, 100:181105, 2012.

\bibitem{Drexhage_1970}
K.~H. Drexhage.
\newblock Influence of a dielectric interface on fluorescence decay time.
\newblock {\em Journal of luminescence}, 1,2:693--701, 1970.

\bibitem{Xiao_2011}
Vamsi~K Komarala and Min Xiao.
\newblock Radiative power of a dipole in the proximity of a dielectric
  interface: a case study of a quantum-dot exciton transition dipole.
\newblock {\em Semicond. Sci. Technol.}, 26:075007, 2011.

\bibitem{Hughes_2009}
Peijun Yao, C.~Van Vlack, A.~Reza, M.~Patterson, M.~M. Dignam, and S.~Hughes.
\newblock Ultrahigh purcell factors and lamb shifts in slow-light metamaterial
  waveguides.
\newblock {\em Physical Review B}, 80:195106, 2009.

\bibitem{Haglund_2001}
B.~J. Lawrie, R.~Mu, and R.~F.~Haglund Jr.
\newblock Substrate dependence of purcell enhancement in ZnO-Ag multilayers.
\newblock {\em Phys. Status Solidi C}, 8:159-162, 2011.

\bibitem{Gresillon_2008}
Emmanuel Fort and Samuel Gresillon.
\newblock Surface enhanced fluorescence (topical review).
\newblock {\em J. Phys. D: Appl. Phys.}, 41:013001, 2008.

\bibitem{Sipe_1984}
J.~M. Wylie and J.~E. Sipe.
\newblock Quantum electrodynamics near an interface.
\newblock {\em Physical Review A}, 30:1185, 1984.

\bibitem{Abdi_2013}
MirFaez Miri, Negar Otrooshi, and Yaser Abdi.
\newblock Nanoemitter in the vicinity of an impedance plane.
\newblock {\em J. Opt. Soc. Am. B}, 30:3027, 2013.

\bibitem{Kunz_JOSA_77}
W.~Lukosz and R.~E. Kunz.
\newblock Light emission by magnetic and electric dipoles close to a plane
  interface. i. total radiated power.
\newblock {\em JOSA}, 67:1607--1615, 1977.

\bibitem{Piche_1992}
Andre Reid and Michel Piche.
\newblock Spontaneous emission in a nonhomogeneous medium: Definition of an
  effective polarizability.
\newblock {\em Phys. Rev. A}, 46:436, 1992.

\end{thebibliography}

\end{document}